\RequirePackage{fix-cm}

\documentclass[twocolumn]{svjour3}          

\def\makeheadbox{{%
\hbox to0pt{\vbox{\baselineskip=10dd\hrule\hbox
to\hsize{\vrule\kern3pt\vbox{\kern3pt
\hbox{This is a pre-print of an article published in the \textbf{Journal of Geodesy}.}
\hbox{The final authenticated version is available online at: \href{https://doi.org/10.1007/s00190-020-01380-w}{https://doi.org/10.1007/s00190-020-01380-w}.}
\kern3pt}\hfil\kern3pt\vrule}\hrule}%
\hss}}}

\smartqed  
\usepackage{graphicx}
  
\usepackage{microtype}

\usepackage{siunitx}
\DeclareSIUnit{\belmilliwatt}{Bm}
\DeclareSIUnit{\dBm}{\deci\belmilliwatt}

\usepackage{xcolor}
\usepackage{authblk}

\usepackage{hyperref}
\hypersetup{colorlinks=true,linkcolor=blue, allcolors=blue}

\usepackage[misc]{ifsym}

\usepackage{natbib}
\usepackage{doi}
\usepackage{amsmath}

\journalname{Journal of Geodesy}
\begin{document}

\title{Methods for coherent optical Doppler orbitography
\thanks{This research was supported by the Australian Research Council's Linkage Infrastructure, Equipment and Facilities (LE160100045) funding scheme; the Australian Research Council's Centre of Excellence for Engineered Quantum Systems (EQUS, CE170100009); and the International Centre for Radio Astronomy Research.}
}

\author{Benjamin P. Dix-Matthews$^{1,2}$\and
        Sascha W. Schediwy$^{1,2}$\and
        David R. Gozzard$^{3}$\and
        Simon Driver$^{1}$\and
        Karl Ulrich Schreiber$^{4}$\and
        Randall Carman$^{5}$\and
        Michael Tobar$^{2}$
}

\authorrunning{B. P. Dix-Matthews et al.} 

\institute{
    \begin{enumerate}
    \item [\Letter] Benjamin P. Dix-Matthews \\ benjamin.dix-matthews@research.uwa.edu.au\\
    \item [$^1$~] International Centre for Radio Astronomy Research, The University of Western Australia, Perth, Australia
    \item [$^2$~] Australian Research Council Centre of Excellence for Engineered Quantum Systems, The University of Western Australia, Perth, Australia
    \item [$^3$~] Department of Quantum Science, Research School of Physics, The Australian National University, Canberra, Australia.
    \item [$^4$~] Research Unit Satellite Geodesy, Technical University of Munich, Munich Germany
    \item [$^5$~] Geoscience Australia, Dongara, Australia
    \end{enumerate}
}

\date{ }

\maketitle

\begin{abstract}
Doppler orbitography uses the Doppler shift in a transmitted signal to determine the orbital parameters of satellites including range and range-rate (or radial velocity). 
We describe two techniques for atmospheric-limited optical Doppler orbitography measurements of range-rate.
The first determines the Doppler shift directly from a heterodyne measurement of the returned optical signal.
The second aims to improve the precision of the first by suppressing atmospheric phase noise imprinted on the transmitted optical signal. 
We demonstrate the performance of each technique over a \SI{2.2}{\kilo\meter} horizontal link with a simulated in-line velocity Doppler shift at the far end. 
A horizontal link of this length has been estimated to exhibit nearly half the total integrated atmospheric turbulence of a vertical link to space.
Without stabilisation of the atmospheric effects, we obtained an estimated range rate precision of \SI{17}{\micro\meter\per\second} at \SI{1}{\second} of integration.
With active suppression of atmospheric phase noise, this improved by three orders-of-magnitude to an estimated range rate precision of \SI{9.0}{\nano\meter\per\second} at \SI{1}{\second} of integration, and \SI{1.1}{\nano\meter\per\second} when integrated over a \SI{60}{\second}. 
This represents four orders-of-magnitude improvement over the typical performance of operational ground to space X-Band systems in terms of range-rate precision at the same integration time.

The performance of this system is a promising proof of concept for coherent optical Doppler orbitography.
There are many additional challenges associated with performing these techniques from ground to space, that were not captured within the preliminary experiments presented here.
In the future, we aim to progress towards a \SI{10}{\kilo\meter} horizontal link to replicate the expected atmospheric turbulence for a ground to space link.

\end{abstract}

\section{Introduction}
Doppler orbitography uses the Doppler shift on a one-way~\citep{Auriol2010} or two-way~\citep{Iess2014} transmission to determine orbital parameters of satellites.
The Doppler Orbitography and Radio positioning Integrated by Satellite (DORIS) system operates at two frequencies (\SI{2036.25}{\mega\hertz} and \SI{401.25}{\mega\hertz}) and is used around the world for geodesy~\citep{Auriol2010}.
This one-way system determines the range rate (or radial velocity) of the satellite at uncertainties of \SI{<0.4}{\milli\meter\per\second}~\citep{MOREAUX2019118}.
Typical operational ground to space X-band radio tracking techniques can now achieve precisions at around \SIrange{20}{100}{\micro\meter\per\second} after \SI{60}{\second} of integration~\citep{Dirkx2018}.

Performing Doppler orbitography measurements in the optical domain offers the potential for several orders of magnitude improvement over traditional microwave techniques.
\citet{chiodo2013lasers} discussed a technique for coherent optical Doppler orbitography based on phase-coherent tracking of an optical signal reflected from a corner cube reflector attached to a satellite in low Earth orbit (LEO).

In 2010, the feasibility of coherent free-space optical transmission was demonstrated over an uncompensated \SI{5}{\kilo\meter} folded horizontal atmospheric link~\citep{djerroud2010coherent}.
The phase noise imprinted on the transmitted signal over this \SI{5}{\kilo\meter} optical link was compared to ground to space turbulence measurements taken from interferometric observations of stars vertically through the atmosphere~\citep{Linfield2001}.
It was found that during calm periods, the \SI{5}{\kilo\meter} link had phase noise similar to the average observed vertically through the atmosphere.
Assuming the atmosphere is approximately reciprocal, this suggests that a two-way ground to space optical Doppler measurement would encounter as much phase noise as a \SI{10}{\kilo\meter} horizontal link.

Active compensation of the free-space optical link to correct for this phase noise~\citep{Gozzard2018,kang2019free} would increase the optical frequency measurement precision of the radial velocity Doppler shift by suppressing noise caused by atmospheric turbulence.
This suggests that an optical stabilisation system capable of operating over a \SI{10}{\kilo\meter} horizontal atmospheric link should be able to improve the quality of range rate measurements when applied to ground to space coherent optical Doppler orbitography.

In this paper, we describe two different techniques for coherent optical Doppler orbitography.
The performance of these techniques correspond to the precision with which the return optical frequency can be measured.
Our first technique determines the Doppler frequency shift directly from a heterodyne measurement of the unstabilised optical return signal, similar to the system designed by~\citet{chiodo2013lasers}.
Our second technique supplements the first with an optical stabilisation system capable of actively correcting the frequency perturbations caused by atmospheric phase noise during transmission, to improve the frequency measurement precision.
We then demonstrate the performance of both of these systems over a \SI{2.2}{\kilo\meter} horizontal link with a simulated Doppler shift at the far end.

This paper is arranged as follows.
In Sect.~\ref{technique} the optical Doppler orbitography techniques are explained.
In Sect.~\ref{method} the system design and the methodology is described.
The results of the demonstration are presented in Sect.~\ref{results}.
Finally, these results and their importance are discussed in Sect.~\ref{discussion}.

\section{Technique}\label{technique}
The focus of this paper is on techniques for turbulence-limited coherent optical Doppler orbitography.
Coherent optical Doppler orbitography involves reflecting a coherent optical beam off a retro-reflector located on a satellite, and using the Doppler shifted return signal to determine the satellite's radial velocity.
The techniques are referred to as `atmospheric turbulence-limited' as it is unable to differentiate the frequency shift caused by the motion of the corner cube from the atmospheric variations operating at the same time scale~\citep{chiodo2013lasers}.
However, in this atmospheric turbulence-limited state, valuable information can be obtained about the dynamics of the turbulent atmosphere~\citep{chiodo2013lasers}.
This is extremely useful information in fields such as coherent optical communications~\citep{Khalighi2014}, and ground-space atomic clock comparisons~\citep{riehle2017optical,lisdat2016clock}.

\begin{figure*}
    \centering
    \includegraphics[width=0.6\linewidth]{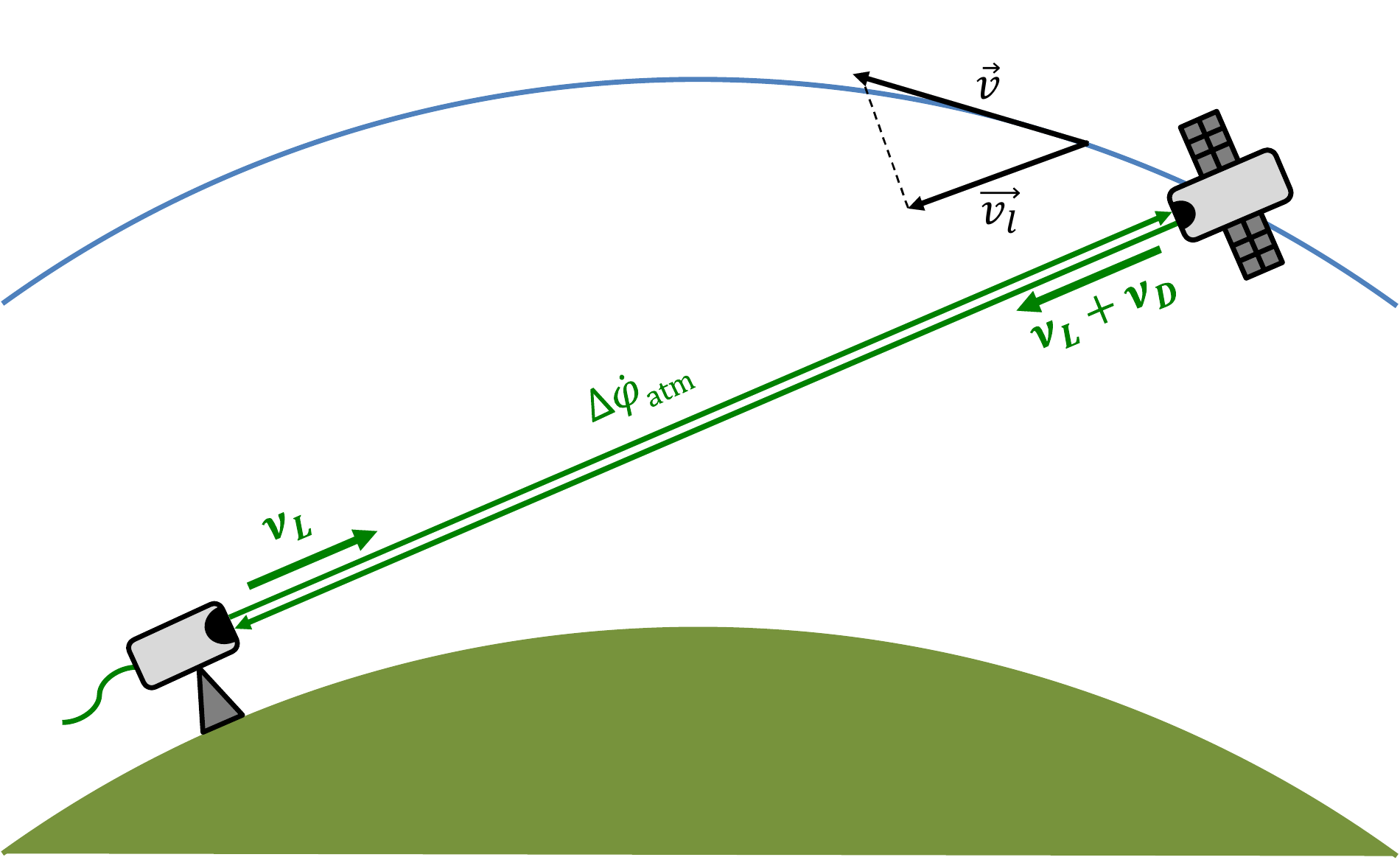}
    \caption{An overview of the technique that is being considered. An optical frequency ($\nu_{L}$) is reflected off a retro-reflector located on the satellite, where it undergoes a Doppler shift ($\nu_{D}$) dependent on the satellite's radial velocity ($\protect\overrightarrow{v_l}$). The satellite's radial velocity is then determined using the Doppler shifted return frequency. The transmitted and reflected signals also experience frequency perturbations caused by the atmospheric phase noise ($\Delta\dot{\phi}_{atm}$).}
    \label{fig:techniqueFigure}
\end{figure*}

We propose two techniques for performing coherent optical Doppler orbitography.
Both techniques utilise an unbalanced Michelson interferometer where the long arm travels over the free-space optical link shown in Fig.~\ref{fig:techniqueFigure} and the short arm provides a local reference.
The signal from the long and short arms are used to make a heterodyne beat which contains the frequency difference between the two.

Our first technique takes the optical Doppler orbitography range rate measurement directly from this frequency difference.
The frequency perturbations caused by the atmospheric phase noise experienced during transmission will remain imprinted on this frequency measurement.

Our second technique supplements the first with an optical frequency stabilisation system that actively suppresses the stochastic atmospheric noise contributions, improving the frequency measurement precision.
The heterodyne beat signal is fed into a servo system that drives the acousto-optic modulator (AOM) in order to compensate for the atmospheric phase noise experienced.
Fluctuations occurring at time scales less than the round trip time of the optical signal cannot be compensated.
This limits the bandwidth of fluctuations that can be corrected to \SI{<300}{\hertz} for LEO objects at \SI{500}{\kilo\meter} when they are directly overhead.
This bandwidth will decrease further for lower angles of elevation.

A major challenge associated with using optical frequency stabilisation techniques for orbitography purposes, is the large frequency shifts caused by the rapidly moving objects in LEO~\citep{chiodo2013lasers}.
These are up to \SI{\pm 10}{\giga\hertz} for objects around \SI{500}{\kilo\meter} when using a near infra-red optical signal (around \SI{193}{\tera\hertz}).
The large frequency deviation results in a high frequency offset between the long and short arms.
This offset will be outside the actuation range of the AOM in the stabilisation system.

This may be overcome through down-conversion of the returned signal using \textit{a priori} knowledge of the satellite's orbit.
Provided the velocity of the orbiting object is known sufficiently accurately, the resultant signal after down-conversion can be brought within the \SIrange{1}{10}{\mega\hertz} bandwidth typical of AOMs.

The resultant signal after down-conversion, will depend on the discrepancy between the predicted radial velocity and true radial velocity, along with any frequency perturbations caused by atmospheric phase noise.
The velocity discrepancy is directly dependent on the orbit of the satellite and thus varies consistently and relatively slowly.
Conversely, we have good reason to believe that atmospheric turbulence will have a zero mean over the timescales in question.
Thus the frequency perturbations caused by atmospheric turbulence can be identified as noise and suppressed.

In the next section, we present an experimental demonstration of how our proposed techniques perform over a \SI{2.2}{\kilo\meter} horizontal link.
The experiment indicates the viability of the technique and provides an estimate of the performance that could be expected in a functioning ground to space coherent optical Doppler orbitography system.

\section{Method}\label{method}
\begin{figure*}
    \centering
    \includegraphics[width=0.8\linewidth]{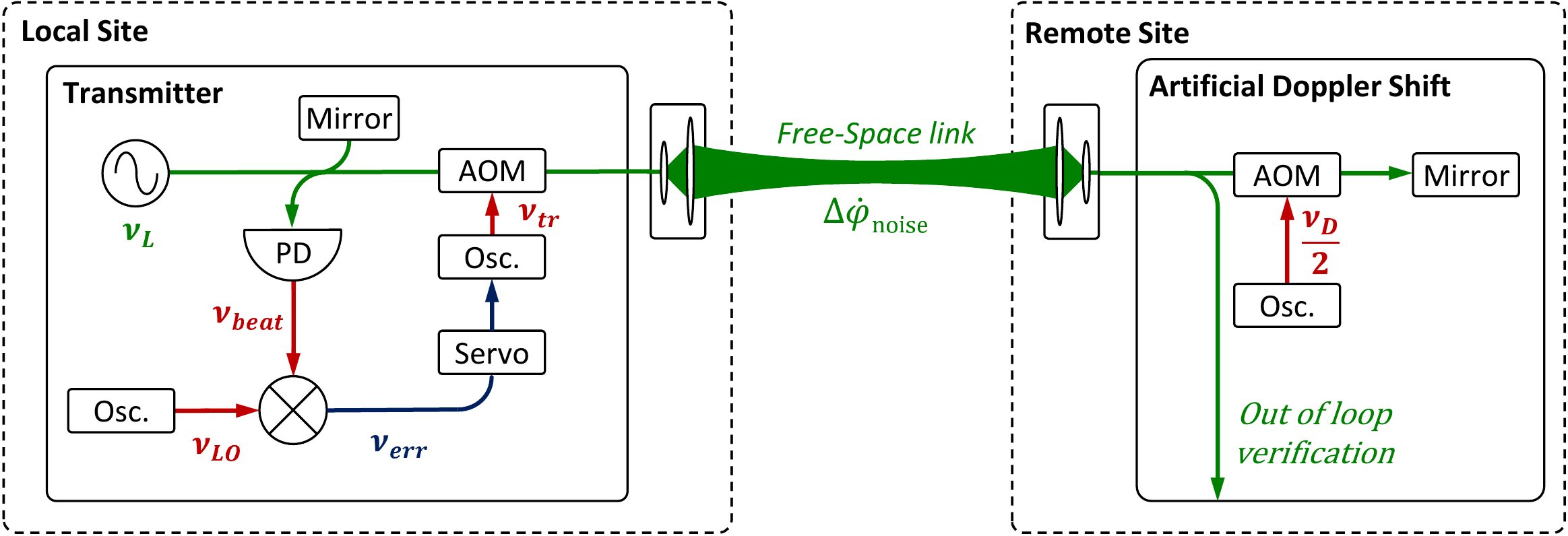}
    \caption{Block diagram of the system used. Green represents optical signals, red represents radio frequency electrical signals, and blue represents the DC electrical control signals.}
    \label{fig:system}
\end{figure*}

The test system we developed passes a highly coherent optical signal with a frequency $\nu_{L}$ (\SI{193}{\tera\hertz}), produced using an NKT Photonics X15 laser, into an unbalanced Michelson interferometer where the short arm is reflected using a Faraday mirror, and the long arm is sent over the free-space link.
Fig.~\ref{fig:system} shows a block diagram of the system used.
At the `local site', the signal is passed through a `transmission' AOM with a nominal frequency ($\nu_{tr}$) of \SI{+50}{\mega\hertz}, which may be varied slightly ($\Delta \nu_{tr}$).
The signal is then launched over the free-space link via a splitter, collimator, and telescope.
The transmission over the free space path adds frequency perturbations ($\Delta\dot{\phi}_{atm}$) to the transmitted signal due to atmospheric turbulence.
The signal reaching the remote terminal ($\nu_{rem}$) has the following frequency.
\begin{equation}\label{remFreq}
    \nu_{rem} = \nu_{L} + \nu_{tr}+ \Delta \nu_{tr}+ \Delta\dot{\phi}_{atm}
\end{equation}

At the remote terminal, the received beam is passed through a \SI{+70}{\mega\hertz} AOM, reflected off a Faraday mirror and sent back through the \SI{+70}{\mega\hertz} AOM.
This \SI{+140}{\mega\hertz} increase ($\nu_{D}$) in the return signal simulates the Doppler shift expected from an object with an radial velocity of \SI{\sim 108}{\meter\per\second}. 

The atmospheric phase noise experienced during transmission over a simple static link, of the type used for these experiment, is approximately reciprocal~\citep{Gozzard2018,Robert2016}.
Frequency perturbations caused by this reciprocal atmospheric phase noise ($\Delta\dot{\phi}_{atm}$) are again added to the signal on the transmission back to the local terminal.
At the local terminal the returning signal is passed back through the transmission AOM again.
The returned signal ($\nu_{rtn}$) now has the following frequency.
\begin{equation}
    \nu_{rtn} = \nu_{L} +2 \nu_{tr}+2 \Delta \nu_{tr}+2  \Delta\dot{\phi}_{atm}+ \nu_{D}
\end{equation}

The return signal $\nu_{rtn}$ is beat against $\nu_{L}$ from the short arm and the optical hetrodyne beat is received by a photodetector (PD).
The resulting electrical beat signal ($\nu_{beat}$) contains information about frequency shift ($\nu_{D}$) (and thus radial velocity) at the far end, as well as a noise component.
\begin{equation}
    \nu_{beat} = 2 \nu_{tr}+2 \Delta \nu_{tr}+2  \Delta\dot{\phi}_{atm}+  \nu_{D}
\end{equation}

For our first technique, the unstabilised Doppler frequency measurement ($\nu_{D_{1}}$) may be obtained through down-conversion by choosing the local oscillator's frequency ($\nu_{LO_{1}}$) according to Eq.~\ref{LOchoice}.
This effectively subtracts the predicted range rate from the return signal.\begin{equation}\label{LOchoice}
    \nu_{LO_{1}} = 2 \nu_{tr}+2 \Delta \nu_{tr}
\end{equation}

The atmospheric effects will remain imprinted on this unstabilised Doppler measurement.
Thus the precision of this measurement will be limited by the total frequency perturbations caused by atmospheric phase noise ($2  \Delta\dot{\phi}_{atm}$).
\begin{equation}
  \nu_{D_{1}} = \nu_{D}+2  \Delta\dot{\phi}_{atm}
\end{equation}

Our second stabilised technique requires that the electrical down-conversion must also remove the predicted Doppler shift before actively suppressing atmospheric phase-noise.
The predicted Doppler shift ($\hat{\nu_{D}}$) is obtained from the \textit{a priori} radial velocity estimate.
\begin{equation}\label{eq:LO2}
    \nu_{LO_{2}} = 2 \nu_{tr}+ \hat{\nu_{D}}
\end{equation}

After this down-conversion, the electrical error signal ($\nu_{err}$) will only contain the turbulence-based frequency perturbations ($2  \Delta\dot{\phi}_{atm}$) and the difference in the expected and real Doppler shift ($\nu_{D} - \hat{\nu_{D}}$).
\begin{equation}
    \nu_{err} =  2 \Delta \nu_{tr}+2  \Delta\dot{\phi}_{atm}+(\nu_{D} - \hat{\nu_{D}}) 
\end{equation}

For our stabilised technique, this error signal is fed into a servo system which varies the frequency of the transmission AOM ($\Delta \nu_{tr}$) in order to actively suppress frequency perturbations caused by the atmospheric phase noise and drive the electrical error signal ($\nu_{err}$) to zero.
The bandwidth of perturbations that can be suppressed is limited by the round trip time of the transmission.
After this suppression, only residual atmospheric frequency perturbations ($2  \Delta\dot{\phi}_{res}$) will remain.\begin{equation}\label{stabilisedAOM}
    \Delta \nu_{tr}= -\Delta\dot{\phi}_{atm}-\frac{(\nu_{D} - \hat{\nu_{D}})}{2}
\end{equation}

The second term of Eq.~\ref{stabilisedAOM} is dependent on the discrepancy between the predicted radial velocity and true radial velocity, and will change gradually over the satellite's pass.
The stabilised Doppler measurement may be obtained from this component of $\Delta \nu_{tr}$.
The precision of the stabilised Doppler measurement will be limited by the residual frequency perturbations.

In our trial case it is assumed that the exact object velocity is known and static, thus $\hat{\nu_{D}}=\nu_{D}=$\SI{140}{\mega\hertz}.
In practice this will vary as the satellite passes, however the situation was simplified for this preliminary experiment.
Therefore, the local oscillator's frequency ($\nu_{LO_{2}}$) in Eq.~\ref{eq:LO2} is \SI{240}{\mega\hertz}.
This results in the return signal being mixed down to a DC error signal.

The performance of each technique was analysed using an analogous system with an out of loop verification.
This system sent the optical signal over a folded path to a corner cube retro-reflector, with the local and remote terminals situated next to each other.

This system used an out of loop heterodyne measurement between the optical signal entering the remote terminal (Eq.~\ref{remFreq}) and the original coherent optical signal ($\nu_{L}$).
The resulting out of loop measurement ($\nu_{OOL}$) after electrical down-conversion by the nominal transmission AOM frequency ($\nu_{tr}$) is given below.
\begin{equation}
    \nu_{OOL} = \Delta \nu_{tr}+ \Delta\dot{\phi}_{atm}
\end{equation}

For the unstabilised technique, the transmission AOM is not used to actively stabilise the link, thus the transmission AOM frequency remains static ($\Delta \nu_{tr} = 0$). 
Thus the unstabilised out of loop measurement ($\nu_{OOL_1}$) will directly measure the one-way frequency perturbations caused by this reciprocal atmospheric phase noise.\begin{equation}
    \nu_{OOL_1}=\Delta\dot{\phi}_{atm}
\end{equation}

The transmission AOM is varied according to Eq.~\ref{stabilisedAOM}.
In this experiment, it is assumed that the Doppler shift is predicted exactly $(\nu_{D} - \hat{\nu_{D}}=0)$.
Thus the out of loop measurement for the second technique ($\nu_{OOL_2}$) will directly measure the residual one-way frequency perturbations after stabilisation.
\begin{equation}
    \nu_{OOL_2}= \Delta\dot{\phi}_{res}
\end{equation}

As the atmosphere is approximately reciprocal~\citep{Gozzard2018,Robert2016}, it was assumed the fractional frequency stability of the two-way transmission ($\frac{\Delta f}{f}$) will be approximately double these one-way measurements ($\frac{\Delta\dot{\phi}_{atm}}{\nu_{L}}$).
The precision of the optical Doppler radial velocity measurements for the unstabilised ($\Delta v_{1}$) and stabilised ($\Delta v_{2}$) systems were estimated using double the fractional frequency stability of the out of loop one-way measurements.
\begin{equation}
\Delta v_{1} \approx \frac{c}{2}\frac{\Delta f}{f} \approx c \frac{\Delta\dot{\phi}_{atm}}{\nu_{L}}
\end{equation}
\begin{equation}
\Delta v_{2} \approx \frac{c}{2}\frac{\Delta f}{f} \approx c \frac{\Delta\dot{\phi}_{res}}{\nu_{L}}
\end{equation}

\subsection{Free-space link}
\begin{figure}
    \centering
    \includegraphics[width=\linewidth]{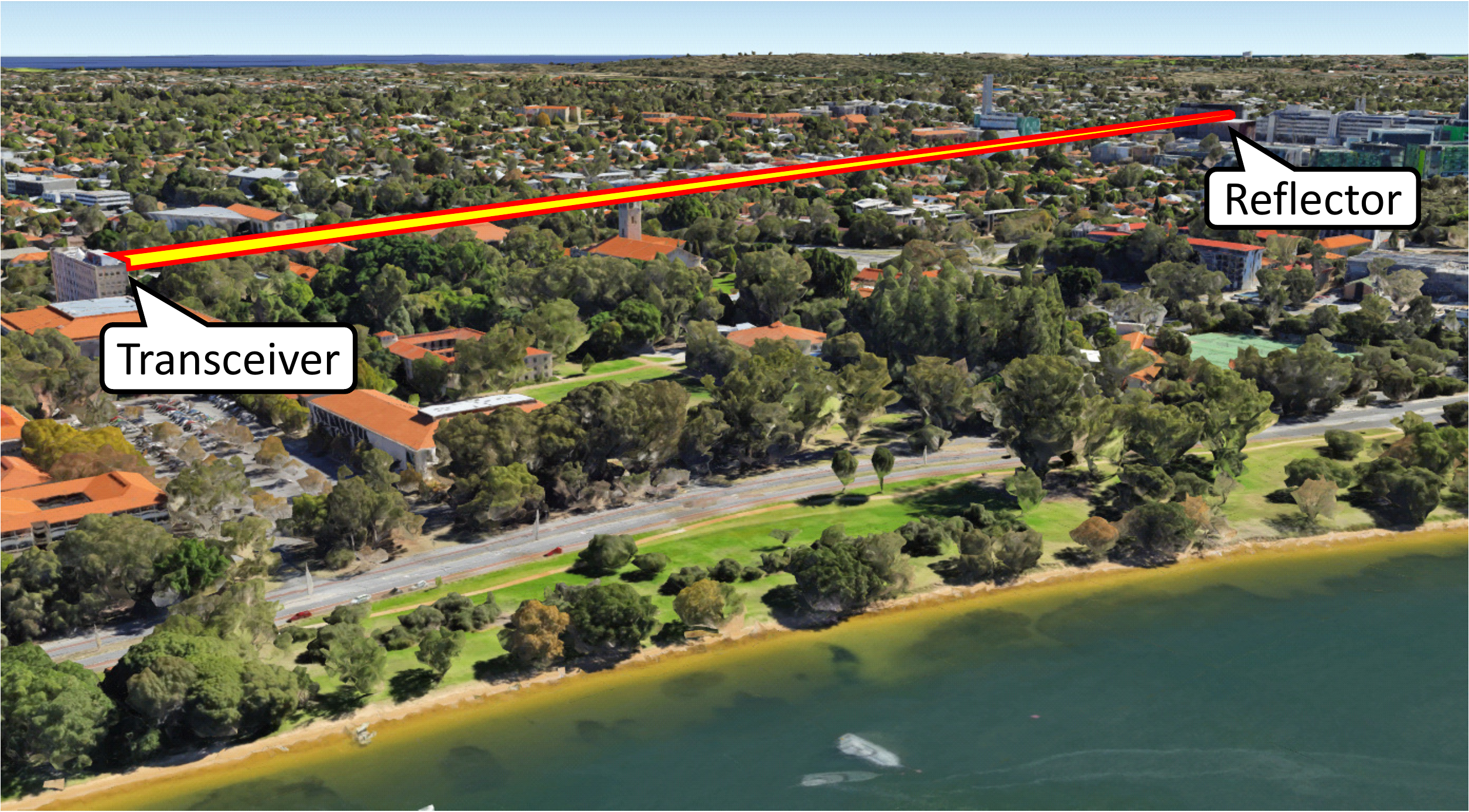}
    \caption{\SI{1.1}{\kilo\meter} free-space optical link from a 5th floor window in the UWA physics building to a 7th floor balcony at HPI. (Modified from Google Maps image)}
    \label{fig:freeSpaceLink}
\end{figure}

The optical terminals and hardware used for transmission over this \SI{2.2}{\kilo\meter} free-space link were identical to those used in our previous paper~\citep{Gozzard2018}.
Detailed discussion of the optical terminals and hardware may be found in our previous paper~\citep{Gozzard2018}.
The physical link used for this experiment reached from the Physics building at the University of Western Australia (UWA) to the Harry Perkins Institute of Medical Research (HPI). 
The main optical terminal and stabilisation hardware were located in a north-facing room, on the 5th floor of the UWA Physics building (approximately \SI{50}{\meter} above sea level). 
This terminal was located behind a closed window, which introduced additional optical loss. 
The corner cube reflector was located \SI{1.1}{\kilo\meter} away, on a south-facing balcony on the 7th floor of HPI. 
As shown in Fig.~\ref{fig:freeSpaceLink}, the link spanned over residential housing and other university buildings.

The optical power output of the transmitter module was approximately \SI{+9}{\dBm} (\SI{7.9}{\milli\watt}).
This was passed through a splitter, collimator and telescope (as shown in~\citet{Gozzard2018}) before being launched over the free-space link with a power of approximately \SI{+2.5}{\dBm} (\SI{1.8}{\milli\watt}).
The free-space link and losses at the corner cube reflector resulted in a net loss of approximately \SI{13}{\deci\bel}.
After passing back through the collimator and splitter, the power of the signal received at the receiver terminal was approximately \SI{-17}{\dBm} (\SI{20}{\micro\watt}).
The remote site has a loss of \SI{11}{\deci\bel}.
After passing back through the splitter, collimator and free-space folded path, the signal received by the transmitter module has a power of approximately \SI{-54}{\dBm} (\SI{4.0}{\nano\watt}).

\section{Results}\label{results}
\begin{figure}
    \label{stabilityResults}
    \centering
    \includegraphics[width=\linewidth]{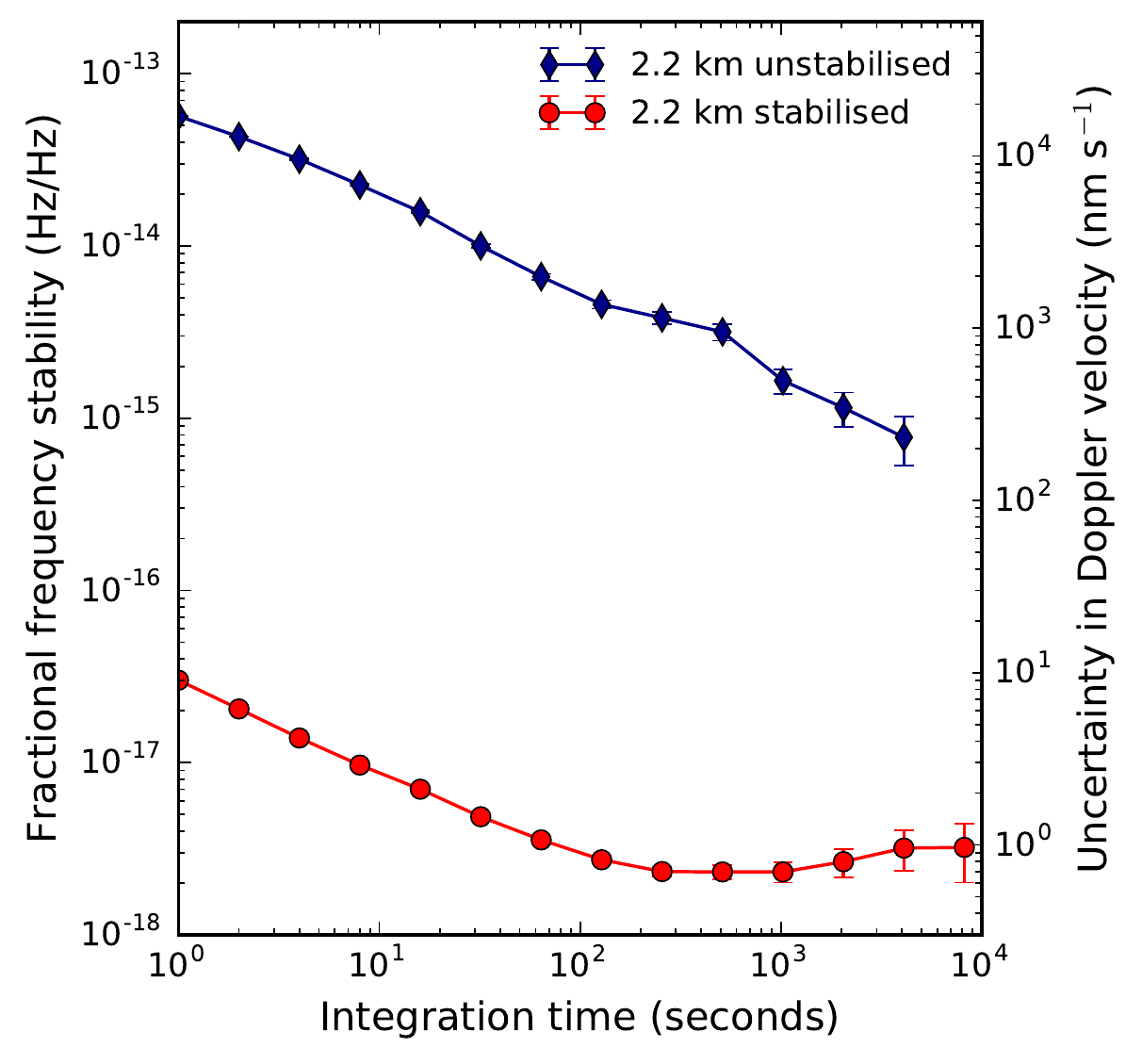}
    \caption{Results for the \SI{2.2}{\kilo\meter} free-space link. Left axis shows the frequency stability and right shows the corresponding uncertainty in Doppler velocity.}
    \label{fig:orbitographyResults}
\end{figure}

An Agilent 53132A $\mathrm{\Lambda}$-type high-precision frequency counter was used to obtain the fractional stability of the optical carrier ($\nu_{L}$) over the \SI{2.2}{\kilo\meter} free-space folded link through an out of loop measurement.
These results are shown in Fig.~\ref{stabilityResults}.

The Agilent 53132A frequency counter averaged the frequency over a \SI{1}{\second} gate time. 
Between each \SI{1}{\second} averaging period, the counter experiences a \SI{0.1}{\second} dead time, which biases the $\tau^{-1}$ slope expected for a white phase noise process to $\tau^{-1/2}$~\citep{Gozzard2018,dawkins2007considerations,faulkner1985time,lesage1983characterization}.

The unstabilised \SI{2.2}{\kilo\meter} free space link, shown with blue diamonds, had a stability of $5.7\times10^{-14}$ at a \SI{1}{\second} integration time ($\tau$). 
At greater integration times this decreased with a $\tau^{-1/2}$ trend, as expected for signals dominated by white phase noise when using a $\mathrm{\Lambda}$-type frequency counter~\citep{Gozzard2018,dawkins2007considerations}.
The fractional stability of the unstabilised \SI{2.2}{\kilo\meter} link corresponded in an uncertainty in Doppler velocity plus atmospheric effects being \SI{17.0}{\micro\meter\per\second} at \SI{1}{\second} of integration. 
This improved to an uncertainty of \SI{2.0}{\micro\meter\per\second} after \SI{60}{\second} of integration.

With active stabilisation, shown with red circles, the fractional frequency stability decreased to $3.0\times10^{-17}$ at a \SI{1}{\second} integration time. 
This three order of magnitude stability improvement over the unstabilised case corresponds to a statistical uncertainty in Doppler velocity of only \SI{9.0}{\nano\meter\per\second} after \SI{1}{\second} of integration time.
This improved down to \SI{1.1}{\nano\meter\per\second} at an integration time of \SI{60}{\second}.
The stability exhibits a $\tau^{-1/2}$ trend up to  about 5 minute integration times, as expected for white phase noise being the dominant noise.
At larger integration times the slope turned upwards, likely due to temperature instability in the measurement room.
The optimum stability reached after about 5 minutes of integration was $2.3\times10^{-18}$, which corresponds to an uncertainty in Doppler velocity of \SI{0.70}{\nano\meter\per\second}.

\section{Discussion}\label{discussion}
The stability performance of the unstabilised version on the \SI{2.2}{\kilo\meter} link shows an order of magnitude improvement over the typical operational performance of ground to space two-way X-band systems at the same integration time~\citep{Dirkx2018}.

When stabilised, this performance improves by more than three orders-of-magnitude and surpasses the performance of other similar proposed optical Doppler ranging systems (currently demonstrated over~\SI{100}{\meter}) by more than an order-of-magnitude~\citep{yang2016high,yang2018optical}.
This optical technique thus holds promise for future Doppler orbitography applications.

Integrating tip-tilt active optics into the system will help to overcome the power fluctuations associated with transmitting through atmospheric turbulence.
When turbulence was at it's minimum, the system could maintain operation on the order of 10 minutes without losing return signal.
Only periods turbulence was at it's minimum were considered when analysing the performance of the two techniques.

It is important to note that the performance of the range rate estimate is referring only to the limit of statistical precision. 
Systematic deviations have not been accounted for in this paper, and so these measurements are therefore not indicative of the actual accuracy of range rate measurements.

\begin{figure}
    \centering
    \includegraphics[width=\linewidth]{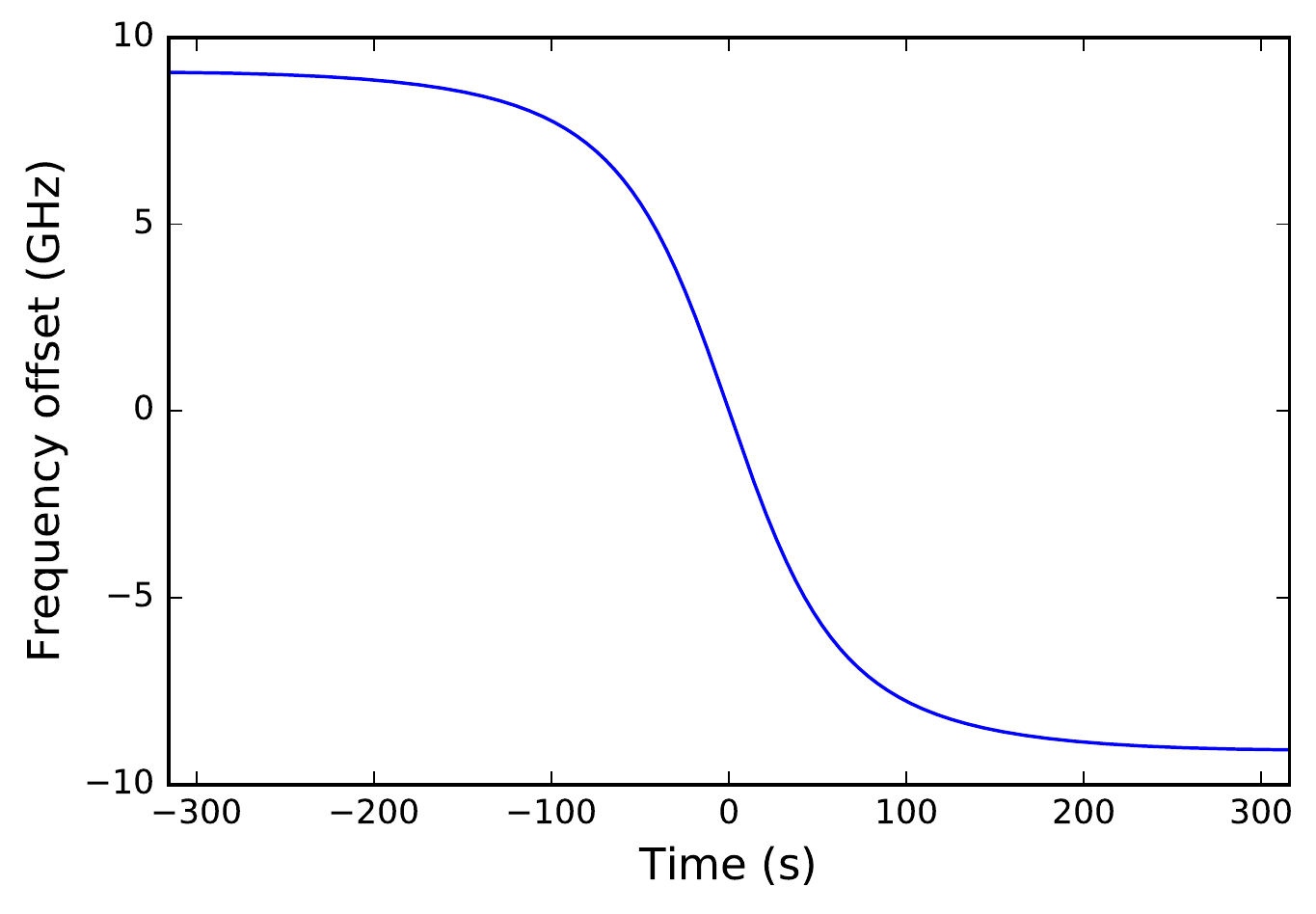}
    \caption{Expected Doppler frequency shift in $\nu_{L}$ (\SI{193}{\tera\hertz}) when reflected off an orbiting object travelling at \SI{7.6}{\kilo\meter\per\second} directly overhead at an altitude of \SI{500}{\kilo\meter}. The orbital pass shown starts with a \SI{20}{\degree} angle of elevation.}
    \label{fig:frequencyOffset}
\end{figure}

The remote site AOM simulated a radial velocity of \SI{108}{\meter\per\second}.
Orbital velocities of objects in LEO are much higher (approximately \SI{7.6}{\kilo\meter\per\second} at an altitude of \SI{500}{\kilo\meter}).
Fig.~\ref{fig:frequencyOffset} shows the expected Doppler frequency shift applied to $\nu_{L}$ (\SI{193}{\tera\hertz}) when reflected off an orbiting object at an altitude of \SI{500}{\kilo\meter}.
It shows one full pass of the object, from horizon to horizon, assuming that the object passes directly overhead.
At these speeds the Doppler frequency shifts span \SI{\pm 10}{\giga\hertz}, with a maximum frequency change rate of \SI{<140}{\mega\hertz\per\second}.

High frequency deviations in the returned signal will undergo down-conversion by mixing the returned signal with a local signal.
The local signal should closely follow the frequency of the returned signal by modelling the transit of the satellite and the corresponding frequency shifts expected.
The local signal used for the down-conversion must therefore be able to sweep the whole frequency range sufficiently quickly to match the expected Doppler frequency shift.
This down-conversion can be conducted optically or electrically.
Optical down-conversion would require a phase locked local laser with a large dynamic range~\citep{chiodo2013lasers}.
Electrical down-conversion could be achieved through the use of microwave electronics and a microwave local oscillator.
As the simulated Doppler shift used in our experiment was a relatively low frequency (\SI{140}{\mega\hertz}), electrical down-conversion was the easier method to implement.

\begin{figure}
    \centering
    \includegraphics[width=\linewidth]{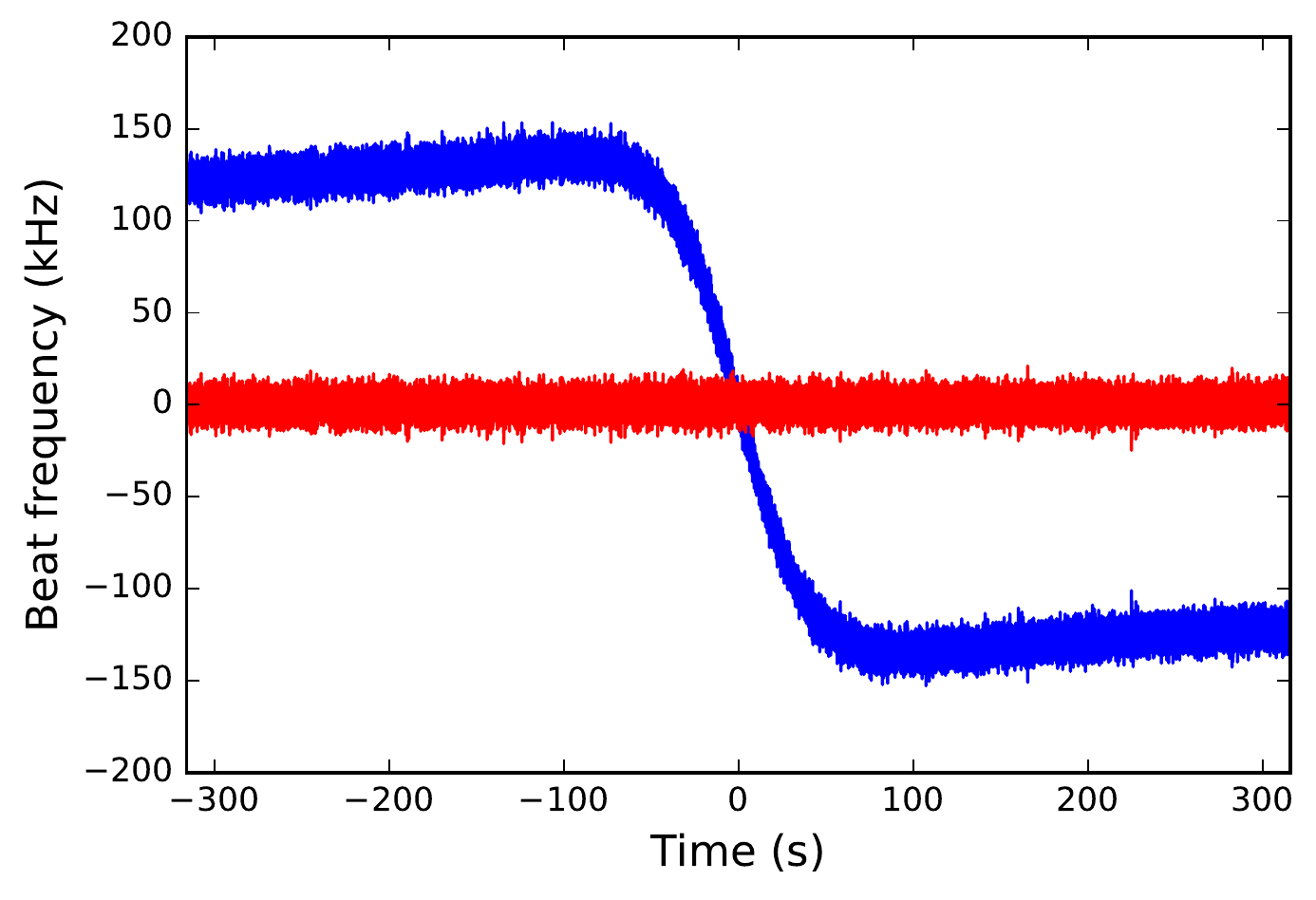}
    \caption{Remaining signal after down-conversion from an object in LEO travelling at \SI{7.6}{\kilo\meter\per\second} directly overhead at an altitude of \SI{500}{\kilo\meter}. Red is the resulting signal assuming no error in the predicted orbiting object's velocity. Blue shows the resulting signal when the predicted speed is \SI{0.1}{\meter\per\second} less than the true speed. Both assume frequency perturbations caused by atmospheric phase noise is an arbitrary white noise process with a mean of \SI{0}{\hertz} and standard deviation of \SI{5}{\kilo\hertz}.}
    \label{fig:downconversionSignal}
\end{figure}

The signal remaining after this down-conversion depends on the accuracy of the satellite velocity estimate and the frequency perturbations caused by atmospheric phase noise.
If the velocity of the satellite were known perfectly, the signal after down-conversion would only have the frequency perturbations caused by atmospheric phase noise during transmission.
This is shown by the red trace in Fig.~\ref{fig:downconversionSignal} for an object travelling at \SI{7.6}{\kilo\meter\per\second},  under the assumption that the frequency perturbations are an arbitrary white noise process with a mean of \SI{0}{\hertz} and standard deviation of \SI{5}{\kilo\hertz}.
This atmospheric frequency perturbations may then be actively suppressed by using the stabilisation system within the transmission AOM bandwidth.

If the estimate of the satellite velocity is incorrect, the resulting signal will have some remaining frequency deviations.
In Fig.~\ref{fig:downconversionSignal}, the remaining frequency offset is caused by modelling the velocity of the orbiting object at a \SI{0.1}{\meter\per\second} less than the true speed.
The velocity estimate (and thus the local signal used for down-conversion) should be varied in order to remove the characteristic frequency deviations remaining.
The components of atmospheric path noise within the bandwidth of the transmission AOM feedback loop may then be actively suppressed using the stabilisation system.

In the next sections, we discuss the many challenges with optical ground to space transmission that have not been captured in this static preliminary test.
It will be vital to address these challenges before the coherent optical Doppler orbitography techniques can be realised.

\subsection{Optical power losses}
The most significant obstacle to performing coherent optical Doppler orbitography appears to be the severe optical power losses encountered during a long-distance ground to space transmission.
Satellite Laser Ranging (SLR) applies a time of flight measurement to obtain the range between a satellite and a ground station. 
This measurement technique can cope with a significant signal loss, well in excess of 15 orders of magnitude~\citep{degnan1993millimeter}. 
In contrast to this, the Doppler measurement concept introduced in this paper requires a higher link margin. 
By choosing a satellite target in the low Earth orbit, a telescope with a larger receive aperture, a tightly focused laser beam and a sufficiently strong laser, successful Doppler observations may be feasible. 
The round trip signal loss, ($L$), can be calculated with the modified link budget Eq.~\ref{linkBudget}, adapted from~\citep{degnan1993millimeter}.
\begin{align}
\begin{split}
    L (dB) ={}& 10 \log \left( \eta_{t} \frac{2}{\pi(\theta_{d}R)^2}e^{-2 \left(\frac{\Delta \theta_{p}}{\theta_{d}}\right)^{2}}\right.\\
    &\times \left[\frac{1}{1+\left(\frac{\Delta \theta_{j}}{\theta_{d}}\right)^{2}}\right] \left.\left( \frac{\sigma A_{r}}{4 \pi R^{2}} \right)\eta_{r}\eta_{q} T_{a}^{2} T_{c}^{2}\right)  
    \label{linkBudget}
    \end{split}
\end{align}

In this equation: $\eta_{t}$ is an efficiency parameter describing the signal loss on the telescope transmit path; $\theta_{d}$ is the Gaussian beam divergence half-angle, which accounts for the focusing of the telescope; $R$ is the distance to the target; $\Delta \theta_{p}$ is the pointing error; $\Delta \theta_{j}$ is the jitter; $\sigma$ represents the Lidar cross section of the retro reflector on the satellite; $A_{r}$ is the effective receive telescope light collecting area; $\eta_{r}$ is an efficiency parameter describing the signal loss on the telescope receive path; $\eta_{q}$ is an efficiency parameter describing the quantum efficiency of the detector; $T_{a}$ is the transmission loss through the atmosphere; and $T_{c}$ is the attenuation caused by a layer of high cirrus clouds.
In these initial calculations, it was assumed that the pointing error ($\Delta \theta_{p}$) and the jitter ($\Delta \theta_{j}$) were negligible.

It is important to note that the losses increase with $R^{4}$, limiting the measurement concept to low Earth orbiting satellites. 
Two promising targets are the GRACE-FO satellites~\citep{fletcher2014}, which are in a low Earth orbit of approximately~\SI{500}{\kilo\meter} with $\sigma$ = 1.8 million square meters. 

We have looked at two different potential ground station candidates from the network of the International Laser Ranging Service~\citep{Pearlman2019}.
One is the Wettzell Laser Ranging System (WLRS), located in Germany, which has the capability to range to the moon.
The system has a larger telescope with a diameter of~\SI{0.75}{\meter} and can tightly focus the beam.
A divergence of only~\SI{1}{\arcsecond} of arc appears to be feasible. 
The other SLR station is located at Yarragadee in Western Australia. 
The receive telescope is also \SI{0.75}{\meter} in diameter and the achievable minimum divergence angle is assumed to be~\SI{3}{\arcsecond} of arc, which may still be slightly optimistic.
For the Yarragadee station, the operation wavelength can only be~\SI{532}{\nano\meter}, while for the WLRS also the fundamental frequency of Nd:YAG of~\SI{1064}{\nano\meter} could be used. 
Table~\ref{tab1} lists typical values for the other parameters of Eq.~\ref{linkBudget} and the corresponding link loss.

\begin{table}
\centering
\caption{Parameter set for the link budget calculation.}
\label{tab1}       
\vspace{0.5cm}

\begin{tabular}{ccc}
\hline\noalign{\smallskip}
Parameter & Wettzell & Yarragadee \\
\noalign{\smallskip}\hline\noalign{\smallskip}
$\eta_{t}$ & 0.6 & 0.75 \\
$\theta_{d}$ & \SI{1}{\arcsecond} & \SI{3}{\arcsecond} \\
$R$ & \SI{500}{\kilo\meter} & \SI{500}{\kilo\meter} \\
$\Delta \theta_{p}$ & $0$ & $0$ \\
$\Delta \theta_{j}$ & $0$ & $0$ \\
$\sigma$ & $1.8\times 10^{6}$m$^{2}$ & $1.8\times 10^{6}$m$^{2}$ \\
$A_r$  & 0.44~m$^{2}$ & 0.44~m$^{2}$\\
$\eta_{r}$ & 0.7 & 0.8 \\
$\eta_{q}$ & 0.5 & 0.5 \\
$T_{a}$ & 0.73 & 0.73 \\
$T_{c}$ & 0.9 & 0.9 \\
\noalign{\smallskip}\hline
Loss &\SI{-86}{\deci\bel} &\SI{-94}{\deci\bel} \\
\end{tabular}
\end{table}

For a high power laser with a transmission power of~\SI{1}{\watt} (\SI{30}{\dBm}), the corresponding received power will be \SI{2.5}{\nano\watt} (\SI{-56}{\dBm}) for Wettzell and \SI{400}{\pico\watt} (\SI{-64}{\dBm}) for Yarragadee.
At this point in time, it seems that dealing with the significant loss over the long distance to the satellite is the largest challenge to be overcome.
High power lasers and weak light tracking techniques~\citep{Francis:14} may be required.

\subsection{Breakdown of link reciprocity}
The point ahead angle required for tracking a real satellite will lead to the forward and return paths passing through slightly different sections of atmosphere which in turn limit accuracy of the assumption of reciprocity that is vital for the stabilisation system~\citep{Robert2016}.
Previous simulation and experimental work found that despite this breakdown in reciprocity, frequency stability better than $2\times10^{-17}$ at a \SI{1}{\second} integration time should be possible for a ground to space link~\citep{Robert2016,swann2019measurement}.
This is below the performance we were able to demonstrate, and thus the breakdown in reciprocity is unlikely to significantly degrade the performance of the techniques presented.

\subsection{Reduction of the servo bandwidth}
At the much longer distance to LEO, the round time of the light will also be greater.
This will limit the bandwidth of the servo loop.
The round trip time will also change dynamically as the distance to the satellite changes.
This is unlikely to be a major issue, as the dynamic reduction in servo bandwidth is largely deterministic, based on \textit{a priori} knowledge of the satellite's orbit.

\subsection{Spacial effects of atmospheric turbulence}
Only the atmospheric phase noise contributions were considered in this preliminary test.
For the real system, spacial variations in the transmitted beam caused by atmospheric turbulence must also be considered~\citep{Robert2016,andrews2005laser,tyson2010principles}.
The integration of either tip-tilt active optics or higher order adaptive optics may be required~\citep{andrews2005laser,tyson2010principles}.
For small apertures, the use of tip-tilt active optics is likely to be sufficient~\citep{Robert2016}.

\subsection{Future development}
Moving forward, the next steps are to increase the optical link towards a \SI{10}{\kilo\meter} horizontal link in order to better capture the challenges associated with severe power losses and atmospheric turbulence.
A higher power fibre laser (\SI{1}{\watt} or \SI{30}{\dBm}) will be used to overcome the additional optical losses and a tip-tilt system will be integrated into our optical terminals in order to combat the increased atmospheric turbulence.

After this horizontal link, we aim to transition towards a more realistic vertical link to an aeronautical object, such as a drone (similar to~\citet{bergeron2019femtosecond}), aeroplane, or balloon.
This will better replicate the Doppler shifts and physical challenges associated with tracking a moving target and be an interim test, before attempting a ground to space transmission at either WLRS or Yarragadee.

\section{Conclusion}
We propose two techniques for performing coherent optical Doppler orbitography measurements on artificial satellites in LEO.
The first performs a heterodyne measurement of the returned optical signal to determine the Doppler shift.
The second technique aims to improve the precision of the first by suppressing atmospheric phase noise imprinted on the transmitted optical signals. 
The performance of these systems were tested over a \SI{2.2}{\kilo\meter} horizontal link.
At the far end the Doppler shift expected from an object with an radial velocity of \SI{\sim 108}{\meter\per\second} was simulated using an AOM.
Without active suppression of the atmospheric phase noise, the estimated range rate precision obtained by our system at a 1 second integration time was \SI{17.0}{\micro\meter\per\second}.
With active phase noise suppression the estimated precision improved by three orders of magnitude to \SI{9.0}{\nano\meter\per\second} at a 1 second integration time.
When integrated over \SI{60}{\second}, this improved to \SI{1.1}{\nano\meter\per\second}.
The stability of this technique on the \SI{2.2}{\kilo\meter} horizontal link represents a 4 order of magnitude improvement in estimated range rate precision over the typical performance of ground to space two-way X-band systems at the same integration time.

This performance is promising; however, other aspects and challenges of ground-to-space laser links must be addressed before using these techniques for ground to space coherent optical Doppler orbitography.
In the future we plan to extend our experiment to better replicate a ground to space laser link through turbulent atmosphere.
This will involve extending the optical link to \SI{10}{\kilo\meter} horizontally and then testing our system over a shorter vertical link to a moving aeronautical object.

\begin{acknowledgements}
We would like to thank Peter Wolf and John Degnan for their helpful discussions.
\end{acknowledgements}

\bibliographystyle{spbasic}      
\small{\bibliography{ref.bib}}

\begin{thebibliography}{26}
\providecommand{\natexlab}[1]{#1}
\providecommand{\url}[1]{{#1}}
\providecommand{\urlprefix}{URL }
\expandafter\ifx\csname urlstyle\endcsname\relax
  \providecommand{\doi}[1]{DOI~\discretionary{}{}{}#1}\else
  \providecommand{\doi}{DOI~\discretionary{}{}{}\begingroup
  \urlstyle{rm}\Url}\fi
\providecommand{\eprint}[2][]{\url{#2}}

\bibitem[{Andrews and Phillips(2005)}]{andrews2005laser}
Andrews LC, Phillips RL (2005) Laser beam propagation through random media, 2nd
  edn. SPIE Press

\bibitem[{Auriol and Tourain(2010)}]{Auriol2010}
Auriol A, Tourain C (2010) {DORIS} system: The new age. Advances in Space
  Research 46(12):1484 -- 1496, \doi{10.1016/j.asr.2010.05.015}

\bibitem[{Bergeron et~al(2019)Bergeron, Sinclair, Swann, Khader, Cossel,
  Cermak, Desch{\^e}nes, and Newbury}]{bergeron2019femtosecond}
Bergeron H, Sinclair LC, Swann WC, Khader I, Cossel KC, Cermak M, Desch{\^e}nes
  JD, Newbury NR (2019) Femtosecond time synchronization of optical clocks off
  of a flying quadcopter. Nature communications 10(1):1819,
  \doi{10.1038/s41467-019-09768-9}

\bibitem[{Chiodo et~al(2013)Chiodo, Djerroud, Acef, Clairon, and
  Wolf}]{chiodo2013lasers}
Chiodo N, Djerroud K, Acef O, Clairon A, Wolf P (2013) Lasers for coherent
  optical satellite links with large dynamics. Appl Opt 52(30):7342--7351,
  \doi{10.1364/AO.52.007342}

\bibitem[{{Dawkins} et~al(2007){Dawkins}, {McFerran}, and
  {Luiten}}]{dawkins2007considerations}
{Dawkins} ST, {McFerran} JJ, {Luiten} AN (2007) Considerations on the
  measurement of the stability of oscillators with frequency counters. IEEE
  Transactions on Ultrasonics, Ferroelectrics, and Frequency Control
  54(5):918--925, \doi{10.1109/TUFFC.2007.337}

\bibitem[{Degnan(1993)}]{degnan1993millimeter}
Degnan JJ (1993) Millimeter accuracy satellite laser ranging: a review.
  Contributions of space geodesy to geodynamics: technology 25:133--162

\bibitem[{Dirkx et~al(2018)Dirkx, Prochazka, Bauer, Visser, Noomen, Gurvits,
  and Vermeersen}]{Dirkx2018}
Dirkx D, Prochazka I, Bauer S, Visser P, Noomen R, Gurvits LI, Vermeersen B
  (2018) {Laser and radio tracking for planetary science missions—a
  comparison}. Journal of Geodesy \doi{10.1007/s00190-018-1171-x}

\bibitem[{{Djerroud} et~al(2010){Djerroud}, {Samain}, {Clairon}, {Acef}, {Man},
  {Lemonde}, and {Wolf}}]{djerroud2010coherent}
{Djerroud} K, {Samain} E, {Clairon} A, {Acef} O, {Man} N, {Lemonde} P, {Wolf} P
  (2010) A coherent optical link through the turbulent atmosphere. In:
  EFTF-2010 24th European Frequency and Time Forum, pp 1--6,
  \doi{10.1109/EFTF.2010.6533653}

\bibitem[{{Faulkner} and {Mestre}(1985)}]{faulkner1985time}
{Faulkner} ND, {Mestre} EVI (1985) Time-domain analysis of frequency stability
  using nonzero dead-time counter techniques. IEEE Transactions on
  Instrumentation and Measurement IM-34(2):144--151,
  \doi{10.1109/TIM.1985.4315289}

\bibitem[{Flechtner et~al(2014)Flechtner, Morton, Watkins, and
  Webb}]{fletcher2014}
Flechtner F, Morton P, Watkins M, Webb F (2014) Status of the {GRACE} follow-on
  mission. In: Marti U (ed) Gravity, Geoid and Height Systems, Springer
  International Publishing, Cham, pp 117--121,
  \doi{10.1007/978-3-319-10837-7_15}

\bibitem[{Francis et~al(2014)Francis, Lam, McKenzie, Sutton, Ward, McClelland,
  and Shaddock}]{Francis:14}
Francis SP, Lam TTY, McKenzie K, Sutton AJ, Ward RL, McClelland DE, Shaddock DA
  (2014) Weak-light phase tracking with a low cycle slip rate. Opt Lett
  39(18):5251--5254, \doi{10.1364/OL.39.005251}

\bibitem[{Gozzard et~al(2018)Gozzard, Schediwy, Stone, Messineo, and
  Tobar}]{Gozzard2018}
Gozzard DR, Schediwy SW, Stone B, Messineo M, Tobar M (2018) Stabilized
  free-space optical frequency transfer. Phys Rev Applied 10:024,046,
  \doi{10.1103/PhysRevApplied.10.024046}

\bibitem[{Iess et~al(2014)Iess, {Di Benedetto}, James, Mercolino, Simone, and
  Tortora}]{Iess2014}
Iess L, {Di Benedetto} M, James N, Mercolino M, Simone L, Tortora P (2014)
  {{ASTRA}: Interdisciplinary study on enhancement of the end-to-end accuracy
  for spacecraft tracking techniques}. Acta Astronautica 94(2):699--707,
  \doi{10.1016/j.actaastro.2013.06.011}

\bibitem[{Kang et~al(2019)Kang, Yang, Chun, Jang, Kim, Kim, and
  Kim}]{kang2019free}
Kang HJ, Yang J, Chun BJ, Jang H, Kim BS, Kim YJ, Kim SW (2019) Free-space
  transfer of comb-rooted optical frequencies over an 18 km open-air link.
  Nature communications 10(1):1--8, \doi{10.1038/s41467-019-12443-8}

\bibitem[{{Khalighi} and {Uysal}(2014)}]{Khalighi2014}
{Khalighi} MA, {Uysal} M (2014) Survey on free space optical communication: A
  communication theory perspective. IEEE Communications Surveys Tutorials
  16(4):2231--2258, \doi{10.1109/COMST.2014.2329501}

\bibitem[{{Lesage}(1983)}]{lesage1983characterization}
{Lesage} P (1983) Characterization of frequency stability: Bias due to the
  juxtaposition of time-interval measurements. IEEE Transactions on
  Instrumentation and Measurement 32(1):204--207,
  \doi{10.1109/TIM.1983.4315042}

\bibitem[{Linfield et~al(2001)Linfield, Colavita, and Lane}]{Linfield2001}
Linfield RP, Colavita MM, Lane BF (2001) {Atmospheric Turbulence Measurements
  with the Palomar Testbed Interferometer}. The Astrophysical Journal
  554(1):505--513, \doi{10.1086/321372}

\bibitem[{Lisdat et~al(2016)Lisdat, Grosche, Quintin, Shi, Raupach, Grebing,
  Nicolodi, Stefani, Al-Masoudi, D{\"o}rscher et~al}]{lisdat2016clock}
Lisdat C, Grosche G, Quintin N, Shi C, Raupach S, Grebing C, Nicolodi D,
  Stefani F, Al-Masoudi A, D{\"o}rscher S, et~al (2016) A clock network for
  geodesy and fundamental science. Nature communications 7:12,443,
  \doi{10.1038/ncomms12443}

\bibitem[{Moreaux et~al(2019)Moreaux, Willis, Lemoine, Zelensky, Couhert,
  Lakbir, and Ferrage}]{MOREAUX2019118}
Moreaux G, Willis P, Lemoine FG, Zelensky NP, Couhert A, Lakbir HA, Ferrage P
  (2019) {DPOD2014}: A new {DORIS} extension of {ITRF2014} for precise orbit
  determination. Advances in Space Research 63(1):118 -- 138,
  \doi{10.1016/j.asr.2018.08.043}

\bibitem[{Pearlman et~al(2019)Pearlman, Noll, Pavlis, Lemoine, Combrink,
  Degnan, Kirchner, and Schreiber}]{Pearlman2019}
Pearlman MR, Noll CE, Pavlis EC, Lemoine FG, Combrink L, Degnan JJ, Kirchner G,
  Schreiber U (2019) The {ILRS}: approaching 20 years and planning for the
  future. Journal of Geodesy \doi{10.1007/s00190-019-01241-1}

\bibitem[{Riehle(2017)}]{riehle2017optical}
Riehle F (2017) Optical clock networks. Nature Photonics 11(1):25,
  \doi{10.1038/nphoton.2016.235}

\bibitem[{Robert et~al(2016)Robert, Conan, and Wolf}]{Robert2016}
Robert C, Conan JM, Wolf P (2016) Impact of turbulence on high-precision
  ground-satellite frequency transfer with two-way coherent optical links. Phys
  Rev A 93:033,860, \doi{10.1103/PhysRevA.93.033860}

\bibitem[{Swann et~al(2019)Swann, Bodine, Khader, Desch\^enes, Baumann,
  Sinclair, and Newbury}]{swann2019measurement}
Swann WC, Bodine MI, Khader I, Desch\^enes JD, Baumann E, Sinclair LC, Newbury
  NR (2019) Measurement of the impact of turbulence anisoplanatism on precision
  free-space optical time transfer. Phys Rev A 99:023,855,
  \doi{10.1103/PhysRevA.99.023855}

\bibitem[{Tyson(2015)}]{tyson2010principles}
Tyson RK (2015) Principles of adaptive optics. CRC press

\bibitem[{{Yang} et~al(2016){Yang}, {Lu}, {Krainak}, and {Sun}}]{yang2016high}
{Yang} G, {Lu} W, {Krainak} M, {Sun} X (2016) High-precision ranging and
  range-rate measurements over free-space-laser communication link. In: 2016
  IEEE Aerospace Conference, pp 1--13, \doi{10.1109/AERO.2016.7500652}

\bibitem[{{Yang} et~al(2018){Yang}, {Chen}, {Numata}, {Krainak}, {Heckler}, and
  {Gramling}}]{yang2018optical}
{Yang} G, {Chen} J, {Numata} K, {Krainak} M, {Heckler} G, {Gramling} C (2018)
  Optical carriers phase based high-precision ranging and range rate
  measurements in coherent optical communication. In: 2018 IEEE Aerospace
  Conference, pp 1--10, \doi{10.1109/AERO.2018.8396734}

\end{thebibliography}

%
%

\end{document}